\documentclass[12pt, table]{article}

\usepackage{graphicx}
\usepackage{authblk}
\usepackage[style=numeric,
            backend=biber,
            sorting=ynt
]{biblatex}
\usepackage{pdfpages} 
\usepackage{array}
\usepackage{longtable}
\usepackage{caption}
\usepackage{csquotes}
\usepackage{booktabs}
\usepackage{booktabs}
\usepackage{tikz}
\usetikzlibrary{shapes.geometric, arrows.meta, positioning}
\usepackage{graphicx}
\usepackage{float}
\usepackage{array}
\usepackage{enumitem}
\usepackage{tabularx}
\usepackage{fullpage}
\usepackage{tcolorbox} 
\usepackage{listings}
\usepackage{amsmath}
\usepackage{graphicx}
\usepackage{lipsum}
\usepackage{awesomebox}
\usepackage{smartdiagram}
    
\definecolor{arimlabs}{rgb}{0.0078, 0.0274, 0.4353}
\usepackage[
    colorlinks=true,
    linkcolor=arimlabs,        
    citecolor=arimlabs,        
    urlcolor=arimlabs       
]{hyperref}

\usetikzlibrary{positioning}
\hypersetup{colorlinks, citecolor=blue}
\usepackage{array}
\usepackage{booktabs}
\usetikzlibrary{positioning, arrows.meta}

\usepackage{listings}
\usepackage{lscape}
\definecolor{codegray}{rgb}{0.95,0.95,0.95}
\definecolor{codered}{rgb}{0.7,0.1,0.1}
\definecolor{codeblue}{rgb}{0.1,0.1,0.7}
\definecolor{codegreen}{rgb}{0,0.5,0.3}

\lstdefinestyle{custompython}{
    language=Python,
    backgroundcolor=\color{codegray},
    basicstyle=\ttfamily\footnotesize,
    keywordstyle=\color{codeblue}\bfseries,
    commentstyle=\color{codegreen}\itshape,
    stringstyle=\color{codered},
    showstringspaces=false,
    tabsize=4,
    breaklines=true,
    frame=single,
    rulecolor=\color{black},
    captionpos=b,
    numbers=none,
    escapeinside={(*@}{@*)}
}

\smartdiagramset{
    border color=none,
    uniform color list=teal!40 for 5 items,
    module x sep=3.75,
    back arrow distance=0.75
}
\title{
\hrule height 1mm
\vspace{25pt}
    Global AI Governance Overview: Understanding Regulatory Requirements Across Global Jurisdictions
\vspace{25pt}
\hrule height 1mm
}
\author[1,2]{Mariia Kyrychenko\textsuperscript{*†}}

\author[1,3]{Mykyta Mudryi\textsuperscript{*}}

\author[1,4]{Markiyan Chaklosh\textsuperscript{*}}
\affil[1]{ARIMLABS.AI}

\affil[2]{University of Texas at Tyler}
\affil[3]{Polish-Japanese Academy of Information Technology}
\affil[4]{University of the National Education Commission in Kraków}
\date{\today}

\affil[ ]{\textit {\{\href{mailto:mmudryi@arimlabs.ai}{mmudryi}, \href{mailto:mchaklosh@arimlabs.ai}{mchaklosh}\}@arimlabs.ai}}

\affil[ ]{\textit{\href{mailto:mariia.kyrychenko@outlook.com}{mariia.kyrychenko@outlook.com}}}

\usepackage[a4paper, top=0.5in, bottom=1in, left=1in, right=1in]{geometry}

\usepackage{fancyhdr}
\begin{document}

\begin{minipage}[h]{\textwidth}
    \maketitle
\begin{abstract}
The rapid development of AI technologies has created a complex regulatory landscape that poses compliance challenges for organizations, yet there remains a lack of clarity about how these requirements translate into practical implementation. This research provides industry professionals with a comprehensive understanding of regulatory requirements across different AI governance frameworks by systematically analyzing over 40 regulatory documents, including the EU AI Act, GDPR, Digital Services Act, US state-level legislation, and Asia-Pacific regulations from China, Japan, and South Korea. Through the analysis, we identify overlapping controls, jurisdiction-specific requirements, and critical implementation gaps that organizations face when translating regulatory obligations into practice. Our analysis reveals four primary challenges: the human oversight paradox where compliance requirements for human control conflict with the autonomous nature of AI systems, insufficient operational guidance for implementing risk assessment frameworks despite detailed structural requirements, absent real-time monitoring systems for training data copyright compliance that rely entirely on reactive enforcement, and fragmented enforcement architecture creating cumulative penalty exposure where single AI decisions can simultaneously violate multiple regulations. Moreover, the research includes a visual regulatory map that illustrates jurisdictional requirements and applicable international frameworks.

\end{abstract}

\end{minipage}

\renewcommand{\thefootnote}{\fnsymbol{footnote}} 
\footnotetext[1]{These authors contributed equally.}
\footnotetext[2]{This work was conducted as part of an internship at ARIMLABS.}

\footnotetext[3]{Correspondence should be addressed to: \href{mailto:research@arimlabs.ai}{research@arimlabs.ai}}
\setcounter{footnote}{1} 

\newpage

\section{Introduction}
The rapid development of AI technologies across industries has created a complex regulatory landscape that poses significant challenges for organizations. As AI systems become increasingly integrated into business operations, industry professionals must navigate an evolving web of compliance requirements that can be difficult to understand and implement effectively. \\

Despite the growing number of AI regulations, laws, and industry standards, there remains a lack of clarity about how these requirements translate into practical implementation. This study is an overview for industry professionals so they are able to incorporate regulatory requirements related to AI, with particular focus on the obligations of specific individuals and leading companies under current regulation.\\

The goal of this research is:
\begin{itemize} 
\item first, to provide industry professionals with a comprehensive understanding of the regulatory requirements across different AI governance frameworks;
\item second, to identify gaps in current regulations.
\end{itemize}
\section{Research Methodology}
The research methodology employed in this study consists of a systematic documentary analysis approach. The documentation that was examined includes AICM Framework \cite{csa2025aicm}, AI Privacy Risks \& Mitigations in LLM \cite{edpb2025privacy}, General Data Protection Regulation (hereinafter GDPR) \cite{gdpr2016}, AI Code of Practice \cite{eu2024codeofpractice}, EU AI Act \cite{euaiact2024}, US AI Action Plan \cite{whitehouse2025aiplan}, NIST AI RMF 1.0 \cite{nist2023airmf}, OWASP Top 10 LLM Vulnerabilities \cite{owasp2025top10}, ISO/IEC 42005:2025 \cite{iso42005:2025}, Digital Services Act (hereinafter DSA) \cite{dsa2022} and AI TRiSM \cite{ibm2024trism}, Directive 2001/29/EC \cite{eucopyright2001consolidated}, Council Regulation (EC) No 1383/2003 \cite{eucustoms2003}, Directive (EU) 2019/790 \cite{eucopyright2019}, California Senate Bill 1001 \cite{california_sb1001_2018}, California Senate Bill 942 \cite{california_sb942_2024}, Utah Senate Bill 149 \cite{utah_sb149_2024}, Arkansas House Bill 1958 \cite{arkansas_hb1958_2025}, Arkansas Act 927 \cite{arkansas_act927_2025}, Maryland House Bill 956 \cite{maryland_hb956_2025}, Montana Senate Bill 212 \cite{montana_sb212_2025}, Illinois House Bill 3773 \cite{illinois_hb3773_2025}, West Virginia House Bill 3187 \cite{westvirginia_hb3187_2025}, China's Cybersecurity Law \cite{creemers2018translation}, China's Personal Information Protection Law \cite{PIPL2021}, Administrative Provisions on Algorithmic Recommendation Services \cite{china2022algorithmic}, Provisions on the Administration of Deep Synthesis Internet Information Services \cite{china2022deepfakes}, Interim Measures for the Management of Generative AI Services \cite{china2023generativeai}, Japan's Social Principles of Human-Centric AI \cite{cabinet2019humancentricAI}, AI Governance Guidelines for Business \cite{meti2024aiguidelinesV1}, AI Governance Framework \cite{grjapan2024aiJapan}, South Korea's Basic Act on AI \cite{korea2024AIBasicAct}, Personal Information Protection Act (PIPA) \cite{koreaPIPA}, Network Act \cite{koreaNetworkAct}, and Product Liability Act \cite{koreaProductLiabilityAct}.\\

Through detailed content analysis of these documents, we examined the specific requirements, obligations, and penalties outlined in each regulatory instrument. Later we identified jurisdiction-specific requirements, and implementation gaps. This qualitative, document-based approach serves the study's dual objectives of providing industry professionals with a comprehensive understanding of regulatory requirements across different AI governance frameworks and identifying significant gaps in current regulations.

\section{Regulatory \& Framework Inventory and Roles}
\subsection{Terminology}
To facilitate the comprehension of the regulations and standards, it is essential to first establish key terminological definitions.\\

\textbf{General-Purpose AI Models (hereinafter GPAI)} are defined by their significant generality and ability to competently execute diverse tasks \cite[Article 3 (63)]{euaiact2024}. These models, often trained with substantial data volumes using self-supervision techniques, are designed for integration across multiple downstream systems and applications.\\

\textbf{Signatories} refer to major AI companies that provide large-scale AI models or AI-powered services and that have committed to voluntary agreements with regulatory bodies, for example entities such as OpenAI and Google \cite{eu2024codeofpractice}.\\

\textbf{Downstream providers} are organizations that integrate general-purpose AI models into their own AI systems or applications \cite{eu2024codeofpractice}. This includes 
entities that: (1) fine-tune or modify GPAI models, (2) build applications that 
incorporate GPAI model functionality beyond simple API access, (3) create AI systems 
where GPAI models perform core functions, or (4) develop products that combine GPAI 
models with additional AI components. Simple API wrappers or user interfaces that 
only facilitate access to unmodified GPAI models without additional AI functionality 
are not considered downstream providers under this definition \cite[Article 3(68)]{euaiact2024}.\\

\textbf{The AI Office} is the European Commission's specialized function responsible for implementing, monitoring, and supervising AI systems and GPAI models, as well as overseeing AI governance \cite[Article 3(47)]{euaiact2024}. All references to the AI Office in the regulation are understood as references to the European Commission itself.\\

\textbf{Deployer} is an entity that uses AI systems within their operations, excluding personal use \cite[Article 3(4)]{euaiact2024}. This includes companies, government agencies, and organizations that implement AI systems in their work.\\

\textbf{Provider} is an organization that develops AI systems or GPAI models and distributes them under their own brand. \cite[Article 3(3)]{euaiact2024}. This includes both direct development and 
commissioning others to develop AI on their behalf.\\

\textbf{Distributor} is an entity that supplies AI systems to the EU market without developing or importing them \cite[Article 3(7)]{euaiact2024}. Examples include software resellers, technology retailers, or platforms that sell third-party AI solutions.\\

\textbf{Systemic risk} means a risk that is specific to the high-impact capabilities of general-purpose AI models, having a significant impact on the European Union market due to their reach, or due to actual or reasonably foreseeable negative effects on public health, safety, public security, fundamental rights, or the society as a whole, that can be propagated at scale across the value chain \cite[Article 3(65)]{euaiact2024}.\\

\textbf{Model Documentation Form} is a form used to describe all the details about how the GPAI works, including all the technical details such as training process, computational resources, and energy consumption \cite{eu2024modeldocform}.

\subsection{EU Regulatory Base}
\subsubsection{Regulation Analysis}

Within the regulatory base of the European Union, four primary legislative documents govern the presence and compliance of AI.\\

In this subsection the analysis of the core regulatory and legal instruments within the EU that describe the obligations and responsibilities of Downstream Providers and Signatories is provided. It includes the EU AI Act, the General Data Protection Regulation (hereinafter GDPR), the AI Code of Practice, and DSA. \\
\begin{itemize}
    \item \textbf{EU AI Act} is a regulation aimed at establishing a framework for the development and use of artificial intelligence on the territory of the European Union \cite{euaiact2024}. The implementation and compliance with EU AI Act is controlled by the AI Office mentioned before.
    \item \textbf{GDPR} is an EU law that sets out rules for how organizations collect, use and protect personal data of individuals within the EU \cite{gdpr2016}. The connection between GDPR and the EU AI Act lies in the fact that GDPR affects the data on which AI systems can be trained, protecting intellectual property and personal identifiable information of individuals.
    \item \textbf{AI Code of Practice} is a voluntary set of guidelines designed to help providers of GPAI models to comply with the EU AI Act \cite{eu2024codeofpractice}. It plays a crucial role in the governance of AI systems.
    \item \textbf{DSA} is a European Union law focused on regulating online platforms and digital services \cite{dsa2022}. Platforms that use AI for content moderation or advertising must ensure that their AI systems meet the standards of both the DSA and EU AI Act.
\end{itemize}
\subsubsection{EU AI Act: Covered Entities}
Based on the analysis of these regulatory documents, this section establishes the scope of compliance obligations under the EU AI Act, identifying which entities are subject to its requirements and which fall outside its purview. \\

Article 53 of EU AI Act states that open-source AI models (specifically GPAI) released under free and open-source licenses do receive exemptions from some AI Act obligations (e.g providing documentation to downstream providers, or upon request, to the AI Office or national authorities), but these exemptions do not apply if the model is classified as having systemic risk \cite[Article 53]{euaiact2024}. In such case, they still must comply with training data summary and copyright policy requirements.
\subsection{EU AI Code of Practice}
The EU AI Code of Practice \cite{eu2024codeofpractice} document is split into three main categories, which are \textbf{Transparency}, \textbf{Copyright}, \textbf{Safety and Security}. This section undertakes a systematic review of the principal implementation guidelines, examining the specific entities and actors to whom these provisions are addressed. While the EU AI Act establishes binding regulatory 
obligations, the Code of Practice provides voluntary guidance to demonstrate compliance with these requirements.
\subsubsection{Transparency Chapter}
The \textbf{Transparency chapter} proposes three measures that serve as guidance for demonstrating compliance with transparency obligations under the AI Act \cite[Article 53, Article 55]{euaiact2024}, specifically addressing Articles 53 and 55 requirements \cite[Transparency Chapter, Objectives]{eu2024codeofpractice}.

\centerline{\textbf{Measure 1.1}  \cite[Transparency Chapter, Measure 1.1]{eu2024codeofpractice}}
Signatories must provide Model Documentation to Downstream Providers and to the AI Office (upon request to assess the compliance with AI Act). It is also required to keep the information up to date and save all the previous versions of the document for a period of \textbf{10 years} after the model has been placed on the market.\\

\centerline{\textbf{Measure 1.2} \cite[Transparency Chapter, Measure 1.2]{eu2024codeofpractice}} 
Signatories must publish or make available the contact details for Downstream Providers and AI Office to request the Model Documentation. The EU AI Code of Practice does not specify the exact format or type of contact details that must be provided. All the required information must be up-to-date and be provided to AI Office in the deadline for the purpose of checking compliance with AI Act. To Downstream Providers the Signatories also must provide the up-to-date information without prejudice within 14 days of receiving the request from them. To the public the Model Documentation can be provided in half, in summary or not provided at all. However, the training content summary must be publicly available.\\

\centerline{\textbf{Measure 1.3 }\cite[Transparency Chapter, Measure 1.3]{eu2024codeofpractice}} 
Signatories must ensure that Model Documentation is controlled for integrity, quality and is protected from unwanted alterations. However, it is important to note that this chapter does not directly affect Downstream Providers, since the Model Documentation Form \cite{eu2024modeldocform} is filled in by Signatories.\\

\subsubsection{Copyright Chapter}
\textbf{Copyright chapter} presents that Signatories should have a copyright policy and controls that correspond to EU Copyright law, which is described in several EU Directives \cite{eucopyright2019} such as Directive 2001/29/EC on the harmonisation of copyright and related rights in the information society \cite{eucopyright2001consolidated} and Directive (EU) 2019/790 on copyright and related rights in the Digital Single Market \cite{eucopyright2019}. Signatories are responsible for verifying that all their measures of copyright are compliant in all the EU member countries. \\

\centerline{\textbf{Measure 1.1 }\cite[Copyright Chapter, Measure 1.1]{eu2024codeofpractice}} 
\vspace{1pt}
Signatories make a copyright policy for the GPAI model placed on EU market, and assign a responsible person who will update and implement this policy within an organization. The copyright policy might be publicly available but not necessary.\\

\centerline{\textbf{Measure 1.2} \cite[Copyright Chapter, Measure 1.2]{eu2024codeofpractice}}
\vspace{1pt}
If the technological denial or other preventing from scraping techniques or signs installed on the web, the Signatories must not try to scrape that information. Also, there is a list with links to resources which must not be touched (for example piracy websites, sites with massive copyright infringement, or the ones that undergo the court process). This list is necessary to help AI companies avoid training their models on content from websites that have been officially recognized by courts or public authorities as systematically violating copyright laws, thereby reducing the risk of incorporating infringing material into their training datasets and subsequently generating copyright-violating outputs. However, this dynamic list has not yet been made publicly available by the European Commission, though it is planned to be published on an EU website \cite{europeancommission2024}.\\
The Signatories must make this information available (on their website for example) and state how their web crawling system works in the technical documentation (Model Documentation Form). The requirements for this form can be found in the Annex IV 2(d) of the EU AI Act \cite[Annex IV]{euaiact2024}.\\

\centerline{\textbf{Measure 1.3} \cite[Copyright Chapter, Measure 1.3]{eu2024codeofpractice}}
\vspace{1pt}

\textbf{Robots.txt Protocol and Technical Restrictions:}\\
The robots.txt protocol should delineate permissible and restricted directories for automated access, and Signatory non-compliance with these access restrictions constitutes a violation of established web crawling conventions and copyright reservation mechanisms. However, if it wasn't specified that a certain directory is not allowed to be accessed, it can be accessed by Signatories. Beyond robots.txt, other technical components and restrictions must be respected.\\

\textbf{Extraterritorial Application:}\\
This requirement is not limited to EU content—the restrictions must be respected everywhere. For example, if OpenAI (US) wanted to train its model on a pirated website operated in Ukraine, it would not be able to sell its services to the EU. The EU AI Act has extraterritorial reach, meaning that non-EU companies must comply with all EU AI Code of Practice requirements including copyright restrictions on training data from any global source, if they wish to access the EU market.\\

\textbf{Communication with Rightsholders:}\\
Signatories are also encouraged to take part in establishing communication with rightsholders (websites from which the data is scraped). Signatories must tell copyright holders how their web crawling system works and keep them informed about any changes.\\

\textbf{Separation of AI Training and Search Functions:}\\
Moreover, Measure 1.3 states that if an AI Model is also a search engine (for example Google), then the two functions must be separated. For example, some website puts 'no AI training' in their metadata, but it still wants to be displayed when the user is looking for information through a search engine. So, Google should not train its Gemini on the data from this website, but still show it to users through the search engine.\\

\centerline{\textbf{Measure 1.4} \cite[Copyright Chapter, Measure 1.4]{eu2024codeofpractice}} 
\vspace{1pt}
Signatories must prevent their models from generating infringing outputs when the downstream provider implements their model. It should be stated in copyright policy, terms, and conditions. Or if it is a model released under open-source license then the prohibition of copyright infringing uses must be stated in the accompanying documentation. 
The infringing rights are defined in Council Regulation (EC) No 1383/2003 of 22 July 2003 \cite{eucustoms2003}. It directly specifies that these are considered infringement:
\begin{itemize}
    \item counterfeit goods;
\item pirated goods;
\item patents;
\item supplementary protection certificates;
\item designs and models;
\item copyright and related rights;
\item trademarks;
\item designations of origin;
\item new plant varieties;
\item geographical indications;
\item any mould or matrix designed or adapted for the manufacture of goods infringing an intellectual property right.
\end{itemize}
While no direct monitoring system exists to detect infringement cases, such incidents must be documented through the Technical Model Documentation, which specifies data collection sources, training datasets, and related information.\\

\centerline{\textbf{Measure 1.5} \cite[Copyright Chapter, Measure 1.5]{eu2024codeofpractice}} 
\vspace{1pt}
Signatories must assign a contact which the rightsholders can contact. Also, there must be a mechanism through which the rightsholders can easily complain to Signatories about copyright infringements. The Signatories must resolve it within the reasonable time frame.\\

The Code of Practice does not specify what exactly must be done with the data (deleted or else). Then , the Directive2001/29/EC must be considered, specifically Article 8 \cite[Article 8]{eucopyright2001consolidated}, which states:\\

“2. Each Member State shall take the measures necessary to ensure
that rightsholders whose interests are affected by an infringing activity
carried out on its territory can bring an action for damages and/or apply
for an injunction and, where appropriate, for the seizure of infringing
material as well as of devices, products or components referred to in
Article 6(2).\\
3. Member States shall ensure that rightsholders are in a position to
apply for an injunction against intermediaries whose services are used
by a third party to infringe a copyright or related right.”\\

According to Article 8 the material must be seizured. Also, Article 17 of GDPR \cite[Article 7]{gdpr2016} states the “right to be forgotten”, which brings a concern with model unlearning methodology.\\

The commitment of data removal does not affect the measures, remedies and sanctions available to enforce copyright and related rights under Union and national law \cite{eu2024codeofpractice}.\\

\subsubsection{Safety and Security Chapter}
The \textbf{Safety and Security} chapter specifies that Signatories must create a framework which is a comprehensive, documented risk management system that establishes organizational policies, procedures, and methodologies for identifying, assessing, and mitigating systemic risks stemming from their AI models throughout the entire lifecycle, implement it, and keep it updated. The requirements for the framework are described in the measures.\\
\centerline{\textbf{Framework requirements}}
\vspace{1pt}
\centerline{\textbf{Measure 1.1} \cite[Safety and Security Chapter, Measure 1.1]{eu2024codeofpractice}}
\vspace{1pt}
Framework must contain the trigger points and their usage. Systemic risk acceptance criteria with tiers and how can it be mitigated. The framework must be confirmed no later than 2 weeks before placing AI model on the market. \\

The term "\textit{trigger points}" in the EU AI Act context refers to computational thresholds that determine when additional regulatory obligations apply to AI models. These are specific thresholds measured in floating-point operations (FLOPs) that trigger different levels of regulatory requirements.\\

\centerline{\textbf{Measure 1.2} \cite[Safety and Security Chapter, Measure 1.2]{eu2024codeofpractice}}
\vspace{1pt}
Signatories must firstly, identify systemic risks, then assess the systemic risks, after that determine whether those risks are acceptable or not and finally implement mitigations to them. The results are reported to AI Office.\\

\centerline{\textbf{Measure 1.3} \cite[Safety and Security Chapter, Measure 1.3]{eu2024codeofpractice}} 
\vspace{1pt}
The framework must stay state-of-art and must be constantly updated. The requirement is to review it every 12 months from the placing of the model on the market, or when it is obviously needed. "\textit{State-of-the-art}" refers to the highest level of development or the most advanced stage of a technology, technique, method, or field of knowledge at a given point in time. It represents the current best practices, most sophisticated capabilities, or cutting-edge achievements available.\\

\centerline{\textbf{Measure 1.4} \cite[Safety and Security Chapter, Measure 1.4]{eu2024codeofpractice}}
\vspace{1pt}
Signatories will provide the AI Office with (unredacted) access to their Framework, and updates thereof, within five business days of either being confirmed.\\

\textbf{\centerline{Systemic Risk Identification}}
\vspace{1pt}
\centerline{\textbf{Measure 2.1} \cite[Safety and Security Chapter, Measure 2.1]{eu2024codeofpractice}} 
\vspace{1pt}
Signatories must identify the systemic risks.\\

\centerline{\textbf{Measure 2.2} \cite[Safety and Security Chapter, Measure 2.2]{eu2024codeofpractice}}
\vspace{1pt}
Signatories must develop systemic risk scenarios including the level of detail for each identified risk.\\

\centerline{\textbf{Systemic Risk Analysis}}
\vspace{1pt}
\centerline{\textbf{Measure 3.1} \cite[Safety and Security Chapter, Measure 3.1]{eu2024codeofpractice}} 
\vspace{1pt}
Signatories must gather model-independent information. For example review the training data, expert interviews, market analysis.\\

\centerline{\textbf{Measure 3.2} \cite[Safety and Security Chapter, Measure 3.2]{eu2024codeofpractice}}
\vspace{1pt}
Signatories must conduct "at least state-of-the-art model evaluations" to assess the model's capabilities, propensities, affordances, and/or effects. The evaluations should use methods like Q\&A sets, benchmarks, red-teaming, etc.\\

\centerline{\textbf{Measure 3.3} \cite[Safety and Security Chapter, Measure 3.3]{eu2024codeofpractice}}
\vspace{1pt}
Signatories must conduct systemic risk modeling.\\

\centerline{\textbf{Measure 3.4} \cite[Safety and Security Chapter, Measure 3.4]{eu2024codeofpractice}}
\vspace{1pt}
Signatories will estimate the probability and severity of harm for the systemic risk.\\

\centerline{\textbf{Measure 3.5} \cite[Safety and Security Chapter, Measure 3.5]{eu2024codeofpractice}}
\vspace{1pt}
Must conduct post-market (collecting user feedback, dialogues with affected stakeholders, collecting information about breaches) monitoring and decide whether the Model Report update is necessary. Signatories must give free access to the model’s version and chain of thought to several independent external evaluators. There are available requirements for the evaluators.\\

\centerline{\textbf{Risk acceptance determination}}
\vspace{1pt}
\centerline{\textbf{Measure 4.1} \cite[Safety and Security Chapter, Measure 4.1]{eu2024codeofpractice}}
\vspace{1pt}
Signatories must identify appropriate systemic risk tiers, and acceptance criteria. Describe how those tiers will be used. \\

\centerline{\textbf{Measure 4.2} \cite[Safety and Security Chapter, Measure 4.2]{eu2024codeofpractice}}
\vspace{1pt}
If the risk acceptance is determined to be not acceptable the Signatories must not make the model available on the market, withdraw, recall it, or restrict. Instead, they have to implement the security mitigations, then conduct one more round of systemic risk identification process.\\

\centerline{\textbf{Safety mitigations}}
\vspace{1pt}
\centerline{\textbf{Measure 5.1} \cite[Safety and Security Chapter, Measure 5.1]{eu2024codeofpractice}}
\vspace{1pt}
Signatories must implement safety mitigations, such as filtering and cleaning training data, access to model, model behaviour, and so on. In the AI EU Code of Practice the following examples of the mitigations are presented:
\begin{itemize}
    \item filtering and cleaning training data, e.g. data that might result in undesirable model propensities such as unfaithful chain-of-thought traces;
 \item monitoring and filtering the model’s inputs and/or outputs;
 \item changing the model behaviour in the interests of safety, such as fine-tuning the model to refuse certain requests or provide unhelpful responses;
 \item staging the access to the model, e.g. by limiting API access to vetted users, gradually expanding access based on post-market monitoring, and/or not making the model parameters publicly available for download initially;
 \item offering tools for other actors to use to mitigate the systemic risks stemming from the model;
 \item techniques that provide high-assurance quantitative safety guarantees concerning the model’s behaviour;
 \item techniques to enable safe ecosystems of AI agents, such as model identifications, specialised communication protocols, or incident monitoring tools; and/or
 \item other emerging safety mitigations, such as for achieving transparency into chain-of-thought reasoning or defending against a model’s ability to subvert its other safety mitigations.
\end{itemize}

\centerline{\textbf{Security mitigations}} 
\vspace{1pt}
\centerline{\textbf{Measure 6.1} \cite[Safety and Security Chapter, Measure 6.1]{eu2024codeofpractice}} 
\vspace{1pt}
Signatories must define a goal that specifies the threat actors including external, insider threats and other.\\

\centerline{\textbf{Measure 6.2} \cite[Safety and Security Chapter, Measure 6.2]{eu2024codeofpractice}}
\vspace{1pt}
Signatories will implement appropriate security mitigations to meet the Security Goal.\\

\centerline{\textbf{Safety and Security Model Reports}}
\vspace{1pt}
\centerline{\textbf{Measure 7.1} \cite[Safety and Security Chapter, Measure 7.1]{eu2024codeofpractice}}
\vspace{1pt}
Signatories must provide a Model Report where they state the deep description of the model. The deep description in the Model Report must contain the following:
\begin{itemize}
    \item a high-level description of the model’s architecture, capabilities, propensities, and affordances, and how the model has been developed, including its training method and data, as well as how these differ from other models they have made available on the market;
\item a description of how the model has been used and is expected to be used, including its use in the development, oversight, and/or evaluation of models;
\item a description of the model versions that are going to be made or are currently made available on the market and/or used, including differences in systemic risk mitigations and systemic risks; and
\item a specification (e.g. via valid hyperlinks) of how Signatories intend the model to operate (often known as a “model specification”), including by:
\begin{itemize}
    \item specifying the principles that the model is intended to follow;
    \item stating how the model is intended to prioritise different kinds of principles and instructions;
    \item listing topics on which the model is intended to refuse instructions; and
    \item providing the system prompt.
\end{itemize}
\end{itemize}

\centerline{\textbf{Measure 7.2} \cite[Safety and Security Chapter, Measure 7.2]{eu2024codeofpractice}}
In the Model Report must be stated the systemic risks and why they are acceptable.\\

\centerline{\textbf{Measure 7.3} \cite[Safety and Security Chapter, Measure 7.3]{eu2024codeofpractice}}
\vspace{1pt}
Also the Model Report must include the results of the systemic risk identification and analysis. As well as the description of all the mitigations implemented.\\

\centerline{\textbf{Measure 7.4} \cite[Safety and Security Chapter, Measure 7.4]{eu2024codeofpractice}}
\vspace{1pt}
In the Model report all he external evaluations must be cited.\\

\centerline{\textbf{Measure 7.5} \cite[Safety and Security Chapter, Measure 7.5]{eu2024codeofpractice}}
\vspace{1pt}
Signatories must ensure that the information about how the development and other things of the model result in the material changes. This must be clear to AI Office.\\

\centerline{\textbf{Measure 7.6} \cite[Safety and Security Chapter, Measure 7.6]{eu2024codeofpractice}} \vspace{1pt}
Model Report updates should be completed within a reasonable amount of time after the Signatory becomes aware of the grounds that necessitate an update, e.g. after discovering them as part of their continuous systemic risk assessment and mitigation. Further, if the model is amongst their respective most capable models available on the market, Signatories will provide the AI Office with an updated Model Report at least every six months. However, Signatories must provide AI Office with the updated model report at least every 6 months, except the cases:
\begin{itemize}
    \item when there were no changes to the model and each identified risk is safe;
    \item when the signatories will place a more capable model on the market within 1 month\\
\end{itemize}
\centerline{\textbf{Measure 7.7} \cite[Safety and Security Chapter, Measure 7.7]{eu2024codeofpractice}}
Signatories provide AI Office with Model Report by the time they place a model on the market, or within 5 days of a confirmed update if the model is updated.\\

\centerline{\textbf{Systemic risk responsibility and allocation}}
\centerline{\textbf{Measure 8.1} \cite[Safety and Security Chapter, Measure 8.1]{eu2024codeofpractice}}
\vspace{1pt}
Signatories must have clear responsibilities. Signatories will clearly define responsibilities for managing the systemic risks stemming from their models across all levels of the organisation.\\

\centerline{\textbf{Measure 8.2} \cite[Safety and Security Chapter, Measure 8.2]{eu2024codeofpractice}}
\vspace{1pt}
The resources must be allocated to the assigned responsible people. The resources are human, financial, computational, information and knowledge.\\

\centerline{\textbf{Measure 8.3} \cite[Safety and Security Chapter, Measure 8.3]{eu2024codeofpractice}}
\vspace{1pt}
Signatories must ensure that people responsible for systemic risk management do it in a healthy way (balanced approach).\\

\centerline{\textbf{Serious incident reporting}}
\centerline{\textbf{Measure 9.1} \cite[Safety and Security Chapter, Measure 9.1]{eu2024codeofpractice}}
Signatories must monitor the threats and incidents through reports. Downstream providers, users, modifiers and other third parties must report any serious incidents to the signatory or the AI office. \\

\centerline{\textbf{Measure 9.2} \cite[Safety and Security Chapter, Measure 9.2]{eu2024codeofpractice}}
\vspace{1pt}
Signatories must report to AI Office with the following information: start and end dates, harm or victim, chain of events, model involved, description, what the Signatory has done or intends to do, what a Signatory recommends the AI Office to do, a root cause analysis, any patterns detected during the post-market monitoring.\\

\centerline{\textbf{Measure 9.3} \cite[Safety and Security Chapter, Measure 9.3]{eu2024codeofpractice}} \vspace{1pt}
Signatories must report if cybersecurity breach happened (no later than 5 days) , or disruption of management/operation (no later than 2 days) , death of a person because of the communication with model ( no later than 10 days) , serious harm to person’s health, or property of the environment (no later than 15 days). The information will be updated by Signatory as available. \\

\centerline{\textbf{Measure 9.4} \cite[Safety and Security Chapter, Measure 9.4]{eu2024codeofpractice}}
\vspace{1pt}
All documentation related to the incident must be kept for 5 years from the breach date or from the documentation date.\\

\centerline{\textbf{Additional documentation and transparency}}

\centerline{\textbf{Measure 10.1} \cite[Safety and Security Chapter, Measure 10.1]{eu2024codeofpractice}}
\vspace{1pt}
Signatories must provide the following documents upon request of AI Office: description of Model’s architecture, how model is integrated into AI systems, model evaluations, safety mitigations implemented. These documents must be retained at least for 10 years after the model has been placed on the market.\\

\centerline{\textbf{Measure 10.2} \cite[Safety and Security Chapter, Measure 10.2]{eu2024codeofpractice}} \vspace{1pt}

Signatories must publish a summarized version of their Framework and Model Reports. If the update is published these documents must also be updated and published, but for frameworks, such publication is not necessary if all of the Signatory’s models are similarly safe or safer models pursuant to Appendix 2.2. For Model Reports, such publication is not necessary if the model is a similarly safe or safer model pursuant to Appendix 2.2.\\

\subsection{EU Penalties}
It is important to note that the Code of Practice itself does not establish or impose any penalties or sanctions. The GPAI Code of Practice is a voluntary tool, prepared by independent experts in a multi-stakeholder process, designed to help industry comply with the AI Act's obligations for providers of general-purpose AI models. The Commission has indicated that it can favourably take adherence to the Code into account when assessing compliance with the AI Act, which may have an effect in deciding upon the amount of regulatory fines. However, all penalty provisions and enforcement mechanisms derive exclusively from the EU AI Act itself, particularly Articles 99-101. The Code of Practice serves only as a compliance demonstration tool and does not create additional legal obligations or sanctions beyond those already established in the AI Act. The penalties are defined by each member state separately, which means that the fines may differ from country to country. Article 99 AI Act (8) \cite[Article 99]{euaiact2024}: Each Member State shall lay down rules on to what extent administrative fines may be imposed on public authorities and bodies established in that Member State.\\

These are the recommendation by EU for Member States:
\begin{itemize}
    \item Non-compliance with the prohibition of the AI practices referred to in Article 5 \cite[Article 5]{euaiact2024}  shall be subject to administrative fines of up to EUR 35 000 000 or, if the offender is an undertaking, up to 7\% of its total worldwide annual turnover for the preceding financial year, whichever is higher \cite[Article 99(3)]{euaiact2024};
    \item Non-compliance with any of the following provisions related to operators or notified bodies, other than those laid down in Article 5 \cite[Article 5]{euaiact2024} , shall be subject to administrative fines of up to EUR 15 000 000 or, if the offender is an undertaking, up to 3\% of its total worldwide annual turnover for the preceding financial year, whichever is higher \cite[Article 99(4)]{euaiact2024};
    \item The supply of incorrect, incomplete or misleading information to notified bodies or national competent authorities in reply to a request shall be subject to administrative fines of up to EUR 7 500 000 or, if the offender is an undertaking, up to 1\% of its total worldwide annual turnover for the preceding financial year, whichever is higher \cite[Article 99(5)]{euaiact2024};
    \item In the case of Small and Medium-sized Enterprises (hereinafter SMEs), including start-ups, each fine referred to in this Article shall be up to the percentages or amount referred to in paragraphs 3, 4 and 5, whichever thereof is lower \cite[Article 99(6)]{euaiact2024}. 
\end{itemize}

Under EU law, \textit{SMEs} are an overarching category of enterprises consisting of three subcategories. Medium-sized enterprises have less than 250 employees and an annual turnover of less than €50 million and/or not more than €43 million on their annual balance sheet. Small enterprises employ less than 50 persons and have an annual turnover and/or balance of less than €10 million. Microenterprises employ less than 10 persons and have an annual turnover and/or balance of less than €2 million.\\

\begin{table}[H]
\centering
\label{tab:sme_classification}
\begin{tabular}{|l|c|c|c|}
\hline
\textbf{Type} & \textbf{Employees} & \textbf{Turnover} & \textbf{Balance Sheet} \\
\hline
\textbf{Medium} & $< 250$ & $\leq$ €50M & $\leq$ €43M \\
\hline
\textbf{Small} & $< 50$ & \multicolumn{2}{c|}{$\leq$ €10M} \\
\hline
\textbf{Micro} & $< 10$ & \multicolumn{2}{c|}{$\leq$ €2M} \\
\hline
\end{tabular}
\caption{EU Classification of Small and Medium-sized Enterprises (SMEs)}
\smallskip
\footnotesize
\end{table}

It is crucial to understand that both Signatories (GPAI model providers) and Downstream Providers face independent penalty exposure under the EU AI Act, though for different violations. GPAI model providers do not face penalties for the Code of Practice itself, as it is a voluntary compliance tool; rather, they face potential fines of up to €35 million or 7\% of global turnover for violating their underlying AI Act obligations under, such as transparency, copyright compliance, and systemic risk assessment requirements. Downstream providers, who integrate these GPAI models into their own AI systems, have separate and independent compliance obligations and can face the same penalty tiers for different violations, including up to €35 million or 7\% of turnover for prohibited AI practices, up to €15 million or 3\% for high-risk AI system violations, and up to €7.5 million or 1\% for providing misleading information. Importantly, Downstream Providers cannot escape liability simply because their GPAI model provider signed the Code of Practice, as the AI Act establishes that regardless of the GPAI provider's compliance status, downstream entities must independently comply with all relevant AI Act requirements for their specific AI systems. This creates a multi-layered accountability structure where both GPAI providers and downstream providers can be penalized simultaneously for their respective violations, ensuring responsibility at every level of the AI value chain.
\subsection{Challenges caused by intersection of GDPR, DSA, and AI Act}

Despite serving different regulatory objectives, the GDPR, DSA, and AI Act establish overlapping compliance requirements that create implementation challenges for organizations operating at the intersection of these frameworks. After establishing the regulatory landscape across jurisdictions, we now provide the implementation challenges that emerged from our analysis of EU requirements.

\begin{enumerate}
\item \textbf{Over-censorship}\\
The implementation of DSA compliance mechanisms through AI content moderation creates an indirect but substantial threat to freedom of expression, particularly for minority voices, as algorithmic bias in automated decision-making leads to the systematic suppression of legitimate speech from underrepresented groups.\\
For example, in 2022, YouTube's AI mistakenly removed a series of videos documenting human rights abuses in Ukraine, classifying them as violent content \cite{barrister2024freespeech}.\\
The intersection of DSA content moderation obligations and AI Act governance requirements creates a systemic over-censorship dynamic, wherein platforms deploy AI systems that prioritize regulatory compliance over speech protection, resulting in algorithmic bias that disproportionately targets minority perspectives while failing to achieve the nuanced content evaluation necessary for protecting fundamental rights.\\
\item \textbf{Technical model explanation}\\
While the EU AI Act's transparency requirements aim to democratize AI understanding through user-friendly model documentation, the technical complexity of neural networks, training algorithms, and data processing pipelines presents challenges for creating explanations that are simultaneously accurate, comprehensive, and accessible to non-technical users. Even though the public disclosure of Model Documentation Report is optional for Signatories, the AI Office and Downstream Providers must still receive this information if they require it, and explanation of the whole model can not be an easy-to-read and concise document. The other questions is, how do AI Office and other authorities make sure that the provided information is correct and the model actually operates the way it is described in the report. \\
\item \textbf{Multiple investigation and fines} \\
A single AI decision can break multiple laws simultaneously, triggering different enforcement mechanisms with potentially cumulative fines. This represents a genuine compliance challenge for platforms using AI systems.
One bad AI decision → User complaint → Platform breaks 3 laws simultaneously → Multiple investigations and fines.
Additional layer of documentation if the companies developing and deploying their own AI moderation tools.
For example, a major social media platform uses an AI system to automatically moderate content, recommend posts, and target advertisements to EU users. The platform's AI system incorrectly flags and removes a post by a journalist from an ethnic minority discussing discrimination issues. The AI also uses this interaction data to profile the user and show them targeted mental health advertisements.
Then the violations would be:

\begin{enumerate}
    \item GDPR \\
The platform's AI system makes automated decisions about content removal without proper human oversight, processes personal data (including sensitive ethnic origin information) unfairly for profiling purposes, and fails to provide transparent information about how the AI processes user data. This would violate Articles 22, 13-14, 5, and 9 of GDPR.
\item DSA\\
The platform fails to properly assess and mitigate systemic risks related to algorithmic bias against minorities, doesn't provide adequate transparency about its recommendation systems, and lacks sufficient notice-and-action procedures for wrongful content removal. In  this case Articles 17, 27, 15, and 34 could be considered as violated.
\item AI Act\\
The AI content moderation system operates as a high-risk AI system without proper risk management, human oversight, or adequate training data governance, leading to discriminatory outcomes. The platform also fails to inform users they're interacting with an AI system, violating transparency obligations. In  this case Articles 13, 14, 16, and 52 could be considered as violated.
\end{enumerate}
When a user files a complaint about this biased AI decision, they unwittingly trigger three separate enforcement mechanisms across different EU authorities, each investigating the same incident under different legal frameworks. Unlike traditional violations that typically fall under a single regulation, AI systems operate at the intersection of multiple regulatory domains, meaning one algorithmic mistake can simultaneously violate data protection, digital services, and AI governance laws.

\item \textbf{Absence of “one-stop shop” for all the regulations} \\
The multi-layered regulation of AI systems creates an accountability paradox: since AI systems simultaneously process personal data, generate content, utilize copyrighted materials, and deploy algorithmic decision-making, violations often implicate multiple regulatory frameworks simultaneously, yet the current institutional architecture assigns different enforcement authorities to each domain, resulting in unclear reporting pathways, potential jurisdictional conflicts, and fragmented oversight that undermines effective enforcement and stakeholder compliance. For example, AI Act violations are reported to the AI Office, DSA breaches to National Digital Services Coordinators, GDPR infractions to Data Protection Authorities, and IP violations to national courts.
\end{enumerate}

\subsection{The US Regulatory Base}
While the EU has established comprehensive enforcement mechanisms, the regulatory landscape in the United States presents a different approach. The US hasn't established a legal framework yet. The president Trump has announced an AI Plan which is directed at "establishing AI dominance in the world". As we can notice, the US' approach differs from EU, because the US is aiming at implementing less restrictions and controls. Certain regulations concerning AI are established only in specific states, with the highest number of laws in California, but on the Federal level there are no regulations yet. As of June 2025, more than half of U.S. states have enacted at least one AI or algorithmic accountability law. In total, 48 states and Puerto Rico have introduced AI-related bills this year, with 26 enacting new measures. Common priorities include transparency, child protection, and reducing algorithmic bias \cite{beckage2025aigovernance}. These laws, including amendments to the California Consumer Privacy Act (CCPA) and specific requirements for AI training data, generative AI (gen AI) disclosures and content labeling, impose new compliance obligations that span a range of sectors, demanding significant operational and technological adjustments.\\
Generally, enforcement from state agencies will come in the form of informal inquiries and formal enforcement actions seeking injunctive relief, fines and, in some cases, criminal penalties. While some laws permit a private right of action, most of the laws focus on state oversight to keep AI transparent and protect the public from misuse. 
\subsubsection{California}
California AI laws apply based on user location, not company registration. Following the same extraterritorial approach as GDPR and CCPA, these laws regulate any company that makes AI systems accessible to California residents, regardless of where the company is incorporated, headquartered, or operates from. For example, \textbf{SB 1001} \cite{california_sb1001_2018} applies to communications with "persons in California," while \textbf{SB 942} \cite{california_sb942_2024} targets AI systems that are "publicly accessible in California" with over 1 million monthly users. This user-centric jurisdictional approach means that any AI company serving California residents, whether based in Silicon Valley, New York, London, or Tokyo, must comply with California's AI regulations, creating a "California Effect" similar to how GDPR applies globally to companies processing EU residents' data \cite{pillsbury2024california}.
\subsubsection{Utah}
Utah's \textbf{SB 149} \cite{utah_sb149_2024} (2024) creates the Artificial Intelligence Policy Act, establishing a comprehensive regulatory framework for AI use in the state. The law establishes liability for AI use that violates consumer protection laws if not properly disclosed, requires clear disclosure when individuals interact with AI in regulated occupations or consumer transactions, and creates the Office of Artificial Intelligence Policy within the Department of Commerce to oversee AI regulation and policy development. The legislation establishes an AI Learning Laboratory Program to assess technologies, risks, and policy implications, while enabling temporary mitigation of regulatory impacts during AI pilot testing through regulatory agreements. The law also clarifies that using AI to commit offenses doesn't exempt someone from criminal responsibility, with enforcement handled by the Division of Consumer Protection through fines up to \$2,500 per violation and court remedies. Notably, the AI Policy Act is set to repeal on May 1, making this essentially a pilot program designed to test regulatory approaches while balancing AI innovation with consumer protection through a "sandbox" framework that allows experimentation under state oversight.
\subsubsection{Arkansas}
Arkansas \textbf{HB 1958} \cite{arkansas_hb1958_2025} requires all public entities in the state to create comprehensive policies regarding the authorized use of artificial intelligence and automated decision tools, with the critical requirement that a human employee or designee always makes the final decision regardless of AI recommendations. The legislation mandates that public entities develop training programs for employees on appropriate AI use, create disciplinary procedures for policy violations, and make these policies publicly available upon request. The policies must include prohibitions on using technology resources for personal political opinions, illegal activities, or circumventing security procedures, establishing a framework for responsible and controlled AI use within Arkansas's public sector while ensuring human oversight remains paramount in government decision-making processes.\\
Arkansas \textbf{Act 927} \cite{arkansas_act927_2025} establishes that individuals who provide prompts or data to generative AI tools will generally own the resulting content or trained model, provided the input data is legally obtained and the content doesn't infringe on existing copyrights or intellectual property rights. The law includes a work-for-hire provision specifying that when employees use generative AI tools as part of their job duties under employer direction and control, the employer owns the resulting content and model training data. This legislation makes Arkansas one of the first U.S. states to define ownership of generative AI output and training results, providing crucial legal clarity for individuals and businesses using AI tools while emphasizing that ownership rights cannot override existing intellectual property protections.
\subsubsection{Maryland}
Maryland \textbf{HB 956} \cite{maryland_hb956_2025} establishes the Workgroup on Artificial Intelligence Implementation to monitor and make recommendations related to AI regulation, consumer protection, current private sector AI use, enforcement authority for the Attorney General's Office of Consumer Protection, and impact on government benefits determination. The workgroup will examine AI's impact on individuals' life opportunities, labor and employment concerns, privacy rights, consumer disclosure standards, and enforcement mechanisms, with required annual reports to legislative committees starting July 1, 2026 \cite{maryland_hb956_2025}. Signed by Governor Wes Moore on April 22, 2025 \cite{maryland_hb956_2025}, the legislation reflects Maryland's proactive approach to studying AI regulation before implementing comprehensive policies, though it faced criticism from consumer advocates who argued the workgroup composition favored industry representatives over civil society voices.
\subsubsection{Montana}
Montana \textbf{SB 212 Right to Compute Act} \cite{montana_sb212_2025} establishes the nation's first "Right to Compute Act," affirming citizens' fundamental right to own, access, and use computational resources including hardware, software, and AI tools under the state constitution's guarantees of property and free expression. The legislation requires deployers of critical infrastructure facilities controlled by AI systems to develop risk management policies based on recognized national and international standards, such as the National Institute of Standards and Technology's AI Risk Management Framework, with annual policy reviews and fallback mechanisms to ensure human control can be restored swiftly. Signed into law by Governor Greg Gianforte on April 17, 2025, the act imposes strict limitations on government restrictions over computational rights, requiring any limitations to be demonstrably necessary and narrowly tailored to fulfill compelling public health or safety interests, while positioning Montana as a leader in protecting individual digital freedoms and technological innovation.
\subsubsection{West Virginia}
West Virginia \textbf{HB 3187} \cite{westvirginia_hb3187_2025} amends the existing AI Task Force established within the Office of the Governor to add identification of economic opportunities to its agenda and require annual reports to the legislature. The task force, composed of state agency heads, legislators, and representatives from AI, cybersecurity, and health industries, is responsible for recommending AI definitions for legislation, assessing AI's impact on workforce and employment, and developing public sector best practices. Approved by the governor on April 25, 2025, and effective July 9, 2025, the legislation extends the task force's termination date to July 1, 2027, positioning West Virginia to systematically explore AI's potential for economic development while ensuring responsible implementation in state government operations.
\subsubsection{Illinois}
Illinois \textbf{HB 3773} \cite{illinois_hb3773_2025} amends the Illinois Human Rights Act to explicitly prohibit employers from using artificial intelligence that subjects employees to discrimination based on protected characteristics, including the specific prohibition against using ZIP codes as a proxy for protected classes. Effective January 1, 2026, the law requires employers to notify employees when AI is used for recruitment, hiring, promotion, training selection, discharge, discipline, tenure, or other employment decisions, though specific notice requirements will be developed by the Illinois Department of Human Rights through future rulemaking. The legislation defines AI broadly as any machine-based system that generates predictions, recommendations, or decisions, including generative AI, and establishes civil rights violations for discriminatory AI use with enforcement through the existing Human Rights Commission process, making Illinois the second state after Colorado to address algorithmic discrimination in employment decisions.

\subsubsection{NIST AI RMF}
The National Institute of Standards and Technology (NIST), a non-regulatory federal agency within the U.S. Department of Commerce, serves as the primary standards-setting body for the United States government and plays a crucial role in developing technological standards and guidelines. While NIST is a U.S. government agency, its frameworks and standards are designed as voluntary guidance that can be adopted by organizations worldwide, carrying no legislative authority or binding legal requirements. In response to growing concerns about artificial intelligence risks and the need for systematic risk management approaches, NIST developed the Artificial Intelligence Risk Management Framework (AI RMF 1.0). This comprehensive framework provides organizations with a structured methodology for identifying, assessing, and mitigating risks associated with AI systems throughout their lifecycle, establishing foundational principles for trustworthy AI development and deployment \cite{nist2023airmf}.\\
The Core functions of NIST AI RMF are \textbf{Govern, Map, Measure, Manage} similar to NIST CSF’s Identify, Protect, Detect, Respond, Recover.\\
\begin{itemize}
    \item \textbf{Govern }function is responsible for putting in place structures, systems, processes, and teams. Also these aspects should be continued and supported all the time.
    \item \textbf{Map} is deciding whether the AI system should be deployed. Does it bring any value? Are the risks mapped? Is the context established and clear? The goal is to understand the context of the AI system.
    \item \textbf{Measure }function is used to quantify and assess the risk. This function uses various metrics and tools to evaluate the AI system's performance, fairness, transparency, and security
    \item \textbf{Manage} is used to allocate the resources to the mapped and measured risks. Basically the goal is to take an action on identified risks.
\end{itemize}
\subsection{Other Frameworks}
\subsubsection{ISO}
Other useful frameworks are brought by International Organization for Standardization (hereinafter ISO) and the key ones that can help govern AI are \textbf{ISO/IEC 42005 \cite{iso42005:2025}, ISO/IEC 42001 \cite{iso42001:2023}, ISO/IEC 23894 \cite{iso23894:2023}}. These ISO standards operate as voluntary international frameworks rather than mandatory federal regulations, providing organizations with structured approaches to AI governance that can be implemented regardless of jurisdiction. These standards offer practical guidance on AI risk management, quality management systems, and ethical considerations that can serve as foundational building blocks for regulatory compliance. For example, for AI Act compliance specifically, these ISO frameworks can be particularly valuable because the EU regulation explicitly recognizes harmonized European standards as a pathway to demonstrate conformity, and ISO standards often inform the development of these harmonized standards. Organizations that proactively implement ISO AI governance frameworks can establish robust documentation, risk assessment procedures, and quality management systems that align with AI Act requirements for transparency, accountability, and risk mitigation, potentially streamlining their compliance efforts and providing evidence of due diligence to regulators across multiple jurisdictions.
Let's consider the difference and main purposes of each of these frameworks:\\
\begin{enumerate}
\item ISO/IEC 42005 – AI Impact Assessment\\
New international standard providing guidance for organizations conducting AI system impact assessments. The document is designed to help organizations evaluate the societal, group, and individual impacts of artificial intelligence technologies—both intended and unintended—across the AI system lifecycle. It helps to answer the question “What does this specific AI system do to the world around?”. Moreover, it contains 4 core methods, which are:\\
\begin{itemize}
    \item Identification: Helping organizations recognize where an AI system may cause harm, create bias, infringe on rights, or produce unintended consequences, including both intended and foreseeable misuses.
    \item Analysis: Offering methods to assess the severity, likelihood, and nature of AI-driven risks, integrating these into broader organizational risk management processes.
    \item Evaluation: Performing the actual assessment, analyzing results, documenting findings, establishing approval workflows.
    \item Monitoring: Implementing continuous monitoring and review mechanisms rather than one-time assessment, with ongoing performance monitoring across diverse populations.
\end{itemize}
\item ISO/IEC 42001 – AI Management\\
Provides an organizational framework, establishing comprehensive governance for all AI activities across the enterprise. It helps an organization to answer the question of “How to govern an AI?”.
\item ISO/IEC 23894 – AI Risk Management\\
This document provides guidance on how organizations can manage risk specifically related to AI, and it is only a guidance, so one can’t get certified in it as it would be with ISO/IEC 27001.\\
\end{enumerate}
The difference between ISO/IEC 42001 and ISO/IEC 42005 is that 42001 is certifiable and provides a framework which is used to govern all AI activities within an organization through policies, procedures and accountable people. Whereas, ISO/IEC 42005 is not certifiable and is a methodology to determine how an individual AI system impacts people, society, and the environment.\\
 ISO/IEC 42001 is \textit{certifiable}, meaning organizations can undergo formal third-party audits to receive official certification that demonstrates their AI management system meets the standard's requirements—similar to how companies can be ISO 9001 certified for quality management or ISO 27001 certified for information security. This certification provides external validation and can be used for compliance demonstrations, procurement requirements, and stakeholder assurance. In contrast, ISO/IEC 42005 is \textit{non-certifiable}, functioning as a guidance document or methodology that organizations can follow internally to assess AI system impacts, but without the option for formal external certification. \\
The timing of ISO/IEC 42005's release aligns perfectly with global regulatory momentum. The European Union's AI Act requires impact assessments for high-risk AI systems. For high-risk AI systems involving personal data processing, organizations must conduct both AI Act conformity assessments and GDPR DPIAs, with the AI Act introducing Fundamental Rights Impact Assessments (FRIAs) that may incorporate DPIA elements to avoid duplication. Organizations must redesign their risk assessment models, integrating DPIA and FRIA into a unified, efficient process, updating templates and workflows to address both privacy and AI-specific aspects.
\subsubsection{NIST AI RMF vs ISO}
Organizations seeking to implement AI governance face a fundamental choice between two primary approaches: the systematic, standards-based methodology offered by the ISO through its AI governance suite (ISO/IEC 42001, 42005, and 23894), or the flexible, framework-oriented approach provided by the U.S. NIST AI RMF 1.0. While both approaches aim to establish trustworthy AI systems through risk management and governance structures, they differ significantly in their implementation requirements, verification mechanisms, and organizational integration methods. The three main differences are:\\
\begin{enumerate}
    \item \textbf{Implementation Flexibility}\\
    The ISO suite mandates specific procedures, formal policies, and standardized documentation requirements that organizations must follow to achieve compliance. In contrast, NIST AI RMF emphasizes achieving risk management outcomes while allowing organizations the flexibility to determine their own implementation methods and adapt the framework to their existing processes and organizational structure.
    \item \textbf{External Validation}\\
    The ISO suite enables formal third-party auditing and certification processes, providing organizations with independent verification of their AI governance systems and the ability to demonstrate compliance through recognized credentials. NIST AI RMF relies on internal self-assessment mechanisms without offering formal external validation pathways, requiring organizations to independently evaluate and document their adherence to the framework's guidance.
    \item \textbf{Legal Recognition}\\
The ISO suite enjoys direct incorporation into binding legislation, with regulations such as the EU AI Act explicitly citing and referencing specific ISO standards as compliance mechanisms. NIST AI RMF operates primarily within US federal policy frameworks and voluntary international guidance documents, lacking the formal legal recognition and mandatory adoption found in hard law regulations across multiple jurisdictions.
\end{enumerate}

\subsubsection{AI TRiSM}
\textbf{AI TRiSM} is another AI Trust, Risk, and Security Management framework created by Gartner and it operates through 4 layers that collectively ensure Ai deployment:
\begin{itemize}
    \item AI Governance establishes organizational-wide policies and accountability structures to oversee all AI initiatives, ensuring alignment with business objectives and ethical standards;
    \item AI Runtime Inspection and Enforcement provides real-time monitoring capabilities to inspect AI models, applications, and agent interactions for anomalies, policy violations, or security threats during operational phases;
    \item Information Governance ensures that AI systems access only properly permissioned and classified data while protecting sensitive information throughout its lifecycle;
    \item Infrastructure and Stack applies traditional technology controls, including endpoint, network, and cloud security solutions, applied to AI workloads.
\end{itemize}
\subsection{Asia-Pacific Regulatory Base}
The Asian AI regulatory landscape presents a diverse and evolving approach to artificial intelligence governance, reflecting the region's position as a global leader in AI development and innovation. While the Asia-Pacific region is expected to become the world's largest AI market by 2030, with countries like China, Japan, Singapore, and South Korea driving significant technological advancement, regulatory approaches vary considerably across jurisdictions \cite{xenoss2025apac}.
\subsubsection{China}
China takes a more hands-on approach to AI regulation. Its government plays a major role in how AI is developed and deployed, primarily through the Interim AI Measures Act (in force since August 15, 2023), zeroing in on risks like disinformation, cyberattacks, discrimination, and privacy breaches.\\
All AI platforms must register AI services, undergo security reviews, label AI-generated content, and ensure data and foundation models come from legitimate, rights-respecting sources. Service providers are also held accountable for content created through their platforms. This framework builds on years of groundwork and ties into broader laws like the \textbf{Personal Information Protection Law} \cite{PIPL2021} and network data security regulations \cite{xenoss2025apac}.\\

\textbf{Administrative Provisions on Algorithmic Recommendation Services \cite{china2022algorithmic}}\\
These provisions are aimed at improving the transparency and accountability of recommendation algorithms used by digital platforms. They require that platforms disclose how their algorithms work, allowing users to understand the basis for content recommendations (e.g., news, videos, products). Key elements include:
\begin{itemize}
    \item User Choice: Users must be able to opt out of personalized recommendations, offering more control over what content they receive.
    \item Algorithm Filing: Platforms must file their algorithmic systems with regulatory authorities for review, ensuring they comply with transparency and fairness standards.
    \item Accountability: Platforms are also required to implement measures that prevent harmful content from being promoted, such as misinformation or content that violates regulations.
\end{itemize}

\textbf{Provisions on the Administration of Deep Synthesis Internet Information Services \cite{china2022deepfakes}}\\
These provisions regulate AI-generated content, often referred to as "deepfakes" or synthetic media. The main points are:

\begin{itemize}
    \item Labeling of AI-Generated Content: Service providers must clearly label content that is generated by AI to prevent deception. This helps consumers identify what is real and what is synthetic.
    \item Misuse Prevention: The provisions aim to stop malicious use of AI-generated content, such as fake news, defamation, or impersonation.
    \item Service Providers' Responsibility: Companies offering deep synthesis services (e.g., voice cloning, AI-generated videos) are required to implement measures to mitigate risks associated with the misuse of their technology.
\end{itemize}

\textbf{Interim Measures for the Management of Generative AI Services \cite{china2023generativeai}}\\
Introduced shortly after the launch of tools like ChatGPT, these interim measures regulate the growing field of generative AI (e.g., text generation, image generation). The key points include:

\begin{itemize}
    \item Data Transparency: Generative AI models must be transparent about their data sources, ensuring that users understand where the AI’s training data comes from.
    \item Accuracy and Accountability: AI services must ensure the accuracy of generated content and are accountable for any misinformation or harmful content generated.
    \item User Protection: The rules emphasize user protection by requiring that generative AI services assess the potential risks before deploying their models, especially regarding privacy and safety.
    \item Security Assessments: A mandatory security review of AI models is required before they are allowed for public deployment, ensuring compliance with national cybersecurity standards.
\end{itemize}

\textbf{Cybersecurity Law (2017) \cite{creemers2018translation}}\\
This landmark law establishes a broad framework for cybersecurity within China, applying to AI services that process sensitive data. Key aspects include:
\begin{itemize}
    \item Data Security: Companies must take appropriate measures to protect the data they collect and process, especially personal or sensitive data.
    \item Security Obligations for AI Services: AI providers must meet cybersecurity standards, including securing user data and preventing attacks on AI infrastructure.
    \item Critical Infrastructure: The law outlines specific regulations for operators of critical information infrastructure, which includes platforms providing AI services.
\end{itemize}

\subsubsection{Japan}
In comparison with China, Japan does not provide any binding regulations, instead it has issued several recommendation documents to promote voluntary compliance and ethical best practices.\\
Key documents include the \textbf{Social Principles of Human-Centric AI (2019) \cite{cabinet2019humancentricAI}}, which outline Japan’s vision for ethical, inclusive, and dignity-centered AI; \textbf{the AI Governance Guidelines for Business \cite{meti2024aiguidelinesV1}}, most recently updated in April 2024, offering practical tools for risk management, transparency, and human oversight; and \textbf{the AI Governance Framework (2024) \cite{grjapan2024aiJapan}}, which consolidates earlier recommendations into a single reference point for voluntary compliance and systematic risk assessment. These soft-law instruments are reinforced by existing statutes such as the Act on the Protection of Personal Information (APPI) — Japan’s GDPR-equivalent — the Digital Platform Transparency Act, which targets fairness in e-commerce and advertising, and the Copyright Act, relevant to generative AI outputs.\\
Institutionally, Japan established the AI Strategy Council as a consultative body, which in May 2024 released a draft discussion paper examining whether binding regulation may become necessary. In parallel, a working group has floated the Basic Act on the Advancement of Responsible AI, a draft bill that, if pursued, would mark a shift from voluntary guidelines to a hard-law regime. The proposal envisions oversight of foundation models, mandatory reporting, and penalties for non-compliance, but it remains in early deliberation.
\subsubsection{South Korea}
South Korea became the first jurisdiction outside the EU to enact a comprehensive AI law, the \textbf{Basic Act on the Development of AI and the Establishment of Trust} (``AI Basic Act''),  officially passed on December 26, 2024, and slated to take effect in January 2026, following a one-year preparation period \cite{korea2024AIBasicAct}. The Act focuses on ``high-impact AI'' systems deployed in critical areas such as healthcare, education, finance, employment, and essential public services. These systems must be designed to allow continuous, meaningful human oversight and intervention, and providers are required to notify users if they are interacting with AI-generated content or decisions, conduct risk assessments, maintain proper documentation, and implement effective control measures \cite{korea2024AIBasicAct}. Additionally, foreign AI providers must appoint a local representative to handle regulatory communications \cite{korea2024AIBasicAct}.

This new law operates alongside existing statutes: the \textbf{Personal Information Protection Act (PIPA)}, which mandates user consent and restricts the scope of personal data collection \cite{koreaPIPA}; the \textbf{Network Act (\textbf{Act on Promotion of Information and Communications Network Utilization and Information Protection}}), covering cybersecurity and data protection for online services \cite{koreaNetworkAct}; and the \textbf{Product Liability Act}, which holds manufacturers accountable for damages arising from defects in AI-driven products, including software faults \cite{koreaProductLiabilityAct}. Together, these regulations establish a robust, multi-layered foundation for trustworthy AI governance in South Korea.

\section{Interpretation Challenges}

Having established the regulatory landscape across EU, US, and Asia-Pacific jurisdictions, we now can identify challenges that emerged from the analysis. While the preceding sections documented what regulations require, this section addresses a question: what challenges do organizations face while translating the obligations into their processes? We identified four challenges that transcend jurisdictional boundaries. These challenges represent not gaps in regulatory coverage, but rather fundamental tensions between legal requirements and technical realities that organizations must navigate independently.

\begin{enumerate}
    \item \textbf{Human Oversight vs. System Autonomy Paradox}
    
There are "inherent concerns that the requirement for human oversight may be fundamentally incompatible with Agentic AI systems, which by definition are designed to act independently to achieve specific goals” \cite{iapp2024agentic}. This creates a fundamental regulatory paradox where compliance requirements conflict with core system capabilities. The EU AI Act Article 14 \cite[Article 14]{euaiact2024} requires that "high-risk AI systems shall be designed and developed in such a way that they can be effectively overseen by natural persons," with oversight measures that are "commensurate with the risks, level of autonomy and context of use". However, the AI Act allows the Commission to consider "the extent to which the AI system acts autonomously" when assessing risk, suggesting that more autonomous systems are more likely to be classified as "high-risk" and thus require more oversight.
\item \textbf{Risk Assessment Framework Implementation Complexity}

While the EU Code of Practice mandates that GPAI providers adopt comprehensive Safety and Security Frameworks with specific structural requirements, including measurable risk tiers, unacceptable risk thresholds, and quantitative risk indicators, it provides insufficient operational guidance on how to translate these high-level requirements into functionally effective risk management systems.\\
The EU AI Act contains the following guidelines on risk measurement:
\begin{itemize}
    \item Article 55(1) - Obligations for Providers of General-Purpose AI  Models with Systemic Risk\\
(a) perform model evaluation in accordance with standardised protocols and tools reflecting the state of the art, including conducting and documenting adversarial testing of the model with a view to identifying and mitigating systemic risks;\\
(b) assess and mitigate possible systemic risks at Union level, including their sources, that may stem from the development, the placing on the market, or the use of general-purpose AI models with systemic risk;

\item Article 51(1) - Classification criteria
"A general-purpose AI model shall be classified as a general-purpose AI model with systemic risk if it meets any of the following conditions: (a) it has high impact capabilities evaluated on the basis of appropriate technical tools and methodologies, including indicators and benchmarks"

\item Article 51(2) - Computational threshold:
"A general-purpose AI model shall be presumed to have high impact capabilities pursuant to paragraph 1, point (a), when the cumulative amount of computation used for its training measured in floating point operations is greater than 10\^ 25."
\end{itemize}
And the EU Code of Practice provides the following guidance:
\begin{itemize}
    \item On Risk Acceptance Criteria\\
"Signatories will describe and justify (in the Framework pursuant to Measure 1.1, point (2)(a)) how they will determine whether the systemic risks stemming from the model are acceptable. To do so, Signatories will: (1) for each identified systemic risk (pursuant to Measure 2.1), at least: (a) define appropriate systemic risk tiers that: (i) are defined in terms of model capabilities, and may additionally incorporate model propensities, risk estimates, and/or other suitable metrics; (ii) are measurable; and (iii) comprise at least one systemic risk tier that has not been reached by the model"
\item On Safety Margins\\
"The safety margin will: (1) be appropriate for the systemic risk; and (2) take into account potential limitations, changes, and uncertainties of: (a) systemic risk sources (e.g. capability improvements after the time of assessment); (b) systemic risk assessments (e.g. under-elicitation of model evaluations or historical accuracy of similar assessments); and (c) the effectiveness of safety and security mitigations (e.g. mitigations being circumvented, deactivated, or subverted)."
\item On Risk Estimation\\
"Signatories shall measure the level of systemic risk stemming from their GPAISR for systemic risks identified pursuant to Commitment II.3, using quantitative and/or qualitative estimates as appropriate for the type and nature of the systemic risk analysed. In doing so, Signatories shall use systemic risk estimation techniques that: are rigorous and state-of-the-art"
\item On Risk Tiers\\
"Such systemic risk acceptance criteria shall: be defined for each of the systemic risks identified as part of systemic risk identification; contain, for at least each of the selected types of systemic risks in Appendix 1.1, systemic risk tiers that: are measurable; are defined in terms of model capabilities, model propensities, harmful outcomes, harmful scenarios, expected harm, quantitative estimates of risk or combinations thereof; and contain at least one systemic risk tier at which that type of systemic risk is considered to be unacceptable"
\end{itemize}
Despite this extensive regulatory text, critical operational questions remain unanswered. The regulations specify what must be measured (systemic risks, model capabilities, harmful outcomes) and that measurements must be "rigorous and state-of-the-art," yet provide no concrete methodology for how organizations should do it.

\item \textbf{Infringement Cases Monitoring}

The regulatory framework relies entirely on post-hoc reporting mechanisms rather than proactive monitoring systems. Enforcement is triggered only when infringement cases are explicitly reported to the AI Office, at which point forensic investigation can commence to verify whether required filtering measures were properly implemented.\\

While the EU Code of Practice, Copyright Chapter, Measure 1.4 \cite{eu2024codeofpractice} requires that Technical Model Documentation specify data collection sources, training datasets, and related information, it provides no corresponding infringement monitoring systems to verify compliance with these documentation requirements.  The way that AI Office can monitor infringement cases is through: \\

\textbf{1. Technical Model Documentation}

The AI Office relies on GPAI providers to self-report their data sources and filtering methodologies within their Technical Model Documentation. This documentation is reviewed upon request or during investigations, but no continuous monitoring validates whether the described systems are actually implemented or whether they effectively prevent prohibited content from entering training datasets.\\

\textbf{2. Reporting Systems}

Infringement cases can only be investigated after they are explicitly reported to the AI Office. Upon receiving a complaint, forensic investigation can commence to verify whether appropriate data usage controls were enforced. However, this reactive approach means violations are addressed only after models have already been trained on potentially prohibited data and potentially deployed to market.\\

No proactive mechanisms exist to detect violations as they occur during the training process, nor to validate that filtering systems described in documentation are functioning as specified.

\end{enumerate}
\section{Conclusion}

Through analysis of over 40 regulatory documents spanning EU, US, and Asia-Pacific jurisdictions, this research provides industry professionals with a clearer picture of what AI governance actually requires in practice. We clarified the regulatory landscape to identify where organizations struggle to bridge the gap between what regulations demand and what's technically feasible.\\

Our analysis revealed four implementation challenges that span across jurisdictions. The \textbf{human oversight paradox} shows a fundamental contradiction: regulations demand human control over systems specifically designed to operate autonomously. Current risk assessment frameworks tell organizations what to measure but offer insufficient guidance on how to measure it, forcing each company to develop their own approach. We found that \textbf{no real-time monitoring systems exist} for training data copyright compliance, enforcement happens only after violations are reported, creating substantial liability exposure for organizations that unknowingly incorporate infringing content. Finally, the fragmented enforcement architecture means a s\textbf{ingle AI decision can simultaneously violate GDPR, DSA, and AI Act provisions}, triggering parallel investigations and potentially cumulative penalties reaching €35 million or 7\% of global turnover.\\

To assist organizations in navigating this complex regulatory environment, Annex \ref{annex:second} provides a visual map of the regulatory landscape examined in this research. The diagram illustrates mandatory requirements across EU, US, and Asia-Pacific jurisdictions, alongside voluntary frameworks and international standards that can support multi-jurisdictional compliance efforts. This visual reference demonstrates how organizations can leverage frameworks such as ISO/IEC 42001, NIST AI RMF, and AI TRiSM to address overlapping requirements across different regulatory regimes.\\

Future research should examine how organizations actually implement these requirements in practice, develop technical solutions for real-time compliance monitoring, and explore new governance models designed specifically for autonomous AI systems. This will require collaboration across disciplines, bringing together policymakers who understand regulatory intent, practitioners facing implementation realities, and researchers developing technical solutions.

\appendix

\includepdf[pages=1,scale=0.9,pagecommand={\section{ Annex}\label{annex:second}}]{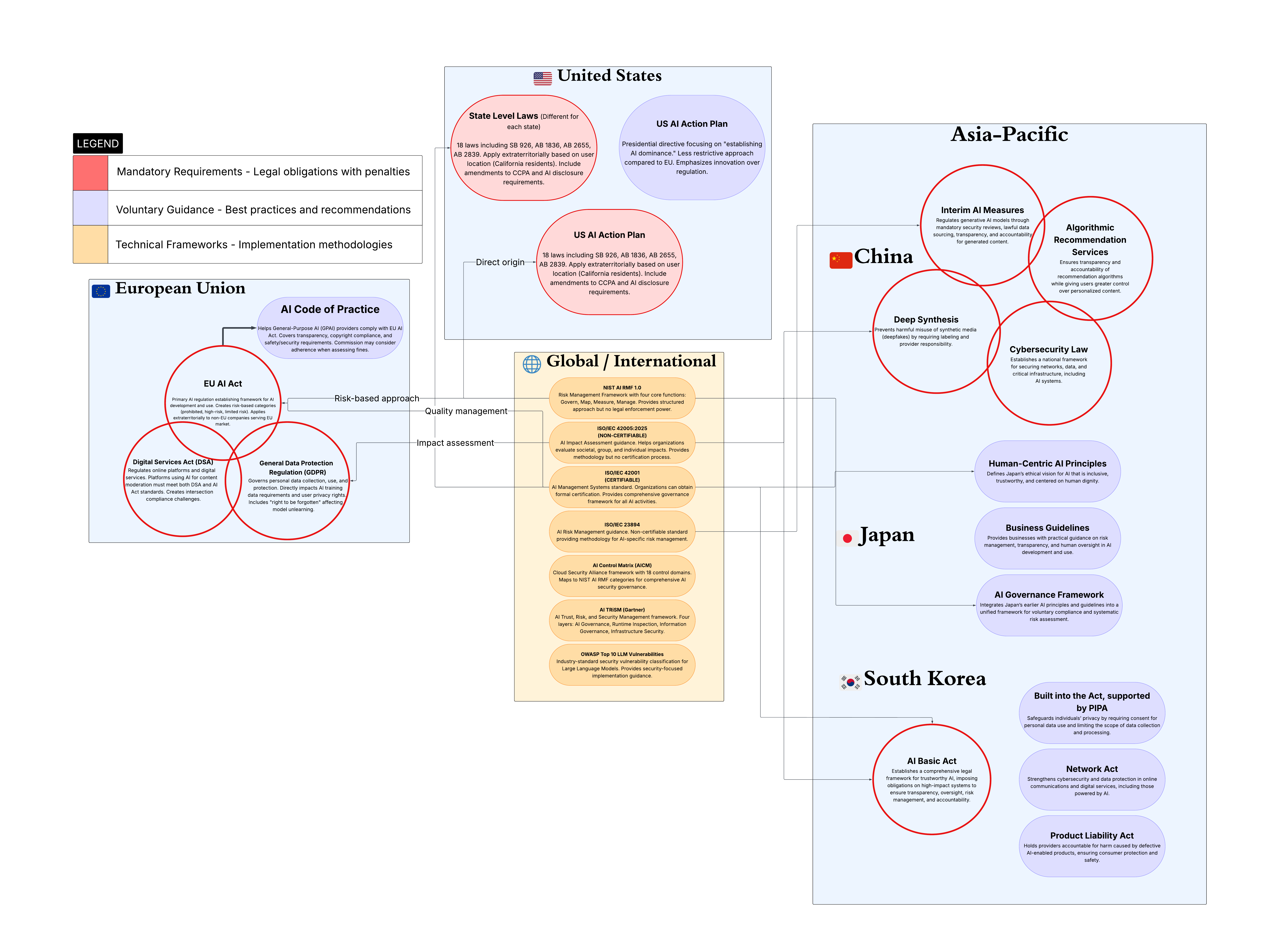}

\section*{\centering Acknowledgement}
The authors would like to thank \textbf{Jakub Łatkiewicz} for his time and effort in reviewing this manuscript and for providing valuable comments and suggestions that helped improve its clarity and quality.
\noindent
\newline
\rule{\textwidth}{0.4pt}

\printbibliography

\end{document}